\documentclass[sigconf,screen,nonacm]{acmart}

\usepackage{amsmath}
\usepackage{subcaption}
\usepackage{graphicx}
\usepackage{placeins}
\usepackage{enumitem}
\usepackage{booktabs}
\usepackage{multirow}
\usepackage{amsfonts}
\usepackage{bm}
\usepackage{upgreek}
\usepackage{xcolor}
\usepackage{tikz}
\usepackage{hyperref}
\usepackage{makecell}

\usetikzlibrary{positioning,fit,arrows.meta,calc}

\def\eqref#1{equation~\ref{#1}}

\def\1{\bm{1}}

\DeclareMathAlphabet{\mathsfit}{\encodingdefault}{\sfdefault}{m}{sl}
\SetMathAlphabet{\mathsfit}{bold}{\encodingdefault}{\sfdefault}{bx}{n}

\hypersetup{
  colorlinks=true,
  linkcolor=blue,
  urlcolor=blue,
  citecolor=blue
}

\acmYear{2026}
\setcctype{by}
\acmConference[ICTIR '26] {Proceedings of the 2026 International ACM SIGIR Conference on Innovative Concepts and Theories in Information Retrieval (ICTIR)}{July 25, 2026}{Melbourne, VIC, Australia.}
\acmBooktitle{Proceedings of the 2026 International ACM SIGIR Conference on Innovative Concepts and Theories in Information Retrieval (ICTIR) (ICTIR '26), July 25, 2026, Melbourne, VIC, Australia}
\acmISBN{979-8-4007-2600-2/2026/07}
\acmDOI{10.1145/3805713.3820413}

\begin{document}

\title{Equity by Design? On the Trade-Offs in Fairness-Driven Recommendation in Heterogeneous Two-Sided Markets}


\author{Dominykas Seputis}
\orcid{0009-0004-9720-1788}

\affiliation{%
  \institution{University of Amsterdam}
  \city{Amsterdam}
  \country{Netherlands}
}

\affiliation{%
  \institution{Vinted}
  \city{Amsterdam}
  \country{Netherlands}
}

\email{d.seputis@uva.nl}
\email{dominykas.seputis@vinted.com}

\author{Alexander Timans}
\orcid{0009-0006-9395-5560}
\affiliation{%
  \institution{University of Amsterdam}
  \city{Amsterdam}
  \country{Netherlands}
}
\email{a.r.timans@uva.nl}

\author{Rajeev Verma}
\orcid{0000-0002-2340-0942}
\affiliation{%
  \institution{University of Amsterdam}
  \city{Amsterdam}
  \country{Netherlands}
}
\email{r.verma@uva.nl}

\renewcommand{\shortauthors}{Dominykas Seputis, Alexander Timans, and Rajeev Verma}

\begin{abstract}
 Two-sided marketplaces embody heterogeneity in incentives: producers seek exposure while consumers seek relevance, and balancing these competing objectives through constrained optimization is now a standard practice. Yet practical platforms face interacting sources of heterogeneity that are often studied separately: multi-item recommendation, heterogeneous consumer groups, and business constraints beyond raw relevance. In this work, we present and study offline optimization framework for analyzing these trade-offs in an unified manner, extending prior two-sided formulations to represent more realistic discrete multi-item recommendations. Within this framework, we couple producer-side exposure guarantees with a consumer-group fairness objective and explicit business-oriented constraints. Our experiments show that the previously reported ``free fairness'' regime from highly stylized single-item recommendation settings disappears once each consumer receives multiple recommendations, and that moderate producer-fairness constraints can improve simulated business metrics by diversifying exposure away from saturated producers. We further show that reduction of inter-group disparity, preserves competitive overall utility.
\end{abstract}




\maketitle

\vspace{5mm}
\section{Introduction}

 Two-sided marketplaces mediate interactions between producers who supply items and consumers who engage with them. Recommendation systems sit at the center of this exchange, determining which producers gain visibility and which consumers receive relevant content. Consider a music streaming platform: artists (producers) want their tracks played, listeners (consumers) want songs they enjoy, and the platform profits when both sides are satisfied. Traditional recommender design optimizes for consumer utility, surfacing the most relevant content under the assumption that satisfied consumers drive engagement \cite{bobadilla2013recommender, li2023recent}. This framing works for single-sided platforms, but in two-sided markets it creates a fundamental tension: producers compete for finite consumer attention, and a system tuned purely for relevance concentrates exposure among a small set of items, marginalizing the rest \cite{singh2018fairness, ye2023seller, 2307.13656}.

\definecolor{hetA}{RGB}{53,106,160}   
\definecolor{hetB}{RGB}{42,130,90}    
\definecolor{hetC}{RGB}{205,103,35}   

\begin{figure*}[t]
\centering
\resizebox{\textwidth}{!}{%
\begin{tikzpicture}[
  font=\footnotesize,
  arr/.style={-{Stealth[length=2.6mm]}, line width=0.8pt, draw=black!60},
  ctag/.style={circle, inner sep=0.9pt, minimum size=11pt, font=\bfseries\scriptsize},
  tagA/.style={ctag, draw=hetA, fill=hetA!12, text=hetA},
  tagB/.style={ctag, draw=hetB, fill=hetB!12, text=hetB},
  tagC/.style={ctag, draw=hetC, fill=hetC!12, text=hetC},
  person/.pic={
    \fill[pic actions] (0,0.085) circle (0.052);
    \fill[pic actions] (-0.092,-0.14) -- (-0.092,-0.05) arc (180:0:0.092) -- (0.092,-0.14) -- cycle;
  },
  note/.pic={
    \fill[pic actions, rotate around={-20:(0,0)}] (0,0) ellipse (0.058 and 0.042);
    \draw[pic actions, line width=0.7pt, line cap=round] (0.052,0.02) -- (0.052,0.235);
    \draw[pic actions, line width=0.7pt, line cap=round] (0.052,0.235) .. controls (0.13,0.20) .. (0.145,0.10);
  },
  goalbox/.pic={
    \draw[pic actions, line width=0.9pt, line join=round] (-0.085,-0.085) rectangle (0.085,0.085);
  },
]

\draw[rounded corners=3pt, draw=black!60, fill=black!2, line width=0.6pt] (0,0.35) rectangle (3.7,5.65);
\node[font=\footnotesize\bfseries] at (1.85,5.34) {Realistic marketplace};
\node[font=\scriptsize, text=black!55] at (1.85,5.0) {(music streaming)};

\node[anchor=west, font=\footnotesize\bfseries, text=black!75] at (0.18,4.64) {Listeners \textnormal{\scriptsize\color{black!55} (seek relevance)}};
\pic[black!60] at (0.40,4.26) {person};
\pic[black!60] at (0.74,4.26) {person};
\pic[black!60] at (1.08,4.26) {person};
\node[anchor=west, font=\footnotesize, text=black!70] at (1.6,4.23) {mainstream};
\pic[black!40] at (0.40,3.84) {person};
\pic[black!40] at (0.74,3.84) {person};
\node[anchor=west, font=\footnotesize, text=black!70] at (1.6,3.81) {casual};
\pic[hetC] at (0.40,3.42) {person};
\node[anchor=west, font=\footnotesize, text=hetC] at (1.6,3.39) {niche};
\node[anchor=west, font=\scriptsize, text=black!55] at (0.18,3.08) {groups $g_1,\dots,g_G$};

\draw[black!25] (0.18,2.88) -- (3.52,2.88);

\node[anchor=west, font=\footnotesize\bfseries, text=black!75] at (0.18,2.62) {Artists \textnormal{\scriptsize\color{black!55} (seek exposure)}};
\pic[black!70, scale=0.9] at (0.36,2.22) {note};
\pic[black!70, scale=0.9] at (0.70,2.22) {note};
\node[anchor=west, font=\footnotesize, text=black!70] at (1.7,2.28) {popular};
\pic[black!35, scale=0.9] at (0.36,1.80) {note};
\pic[black!35, scale=0.9] at (0.70,1.80) {note};
\pic[black!35, scale=0.9] at (1.04,1.80) {note};
\pic[black!35, scale=0.9] at (1.38,1.80) {note};
\node[anchor=west, font=\footnotesize, text=black!70] at (1.7,1.86) {long-tail};
\node[anchor=west, font=\scriptsize, text=black!55] at (0.18,1.50) {values $v_j$ (price, margin)};

\draw[black!25] (0.18,1.3) -- (3.52,1.3);

\foreach \col/\rowa/\rowb/\rowc in {0/15/55/30, 1/60/20/10, 2/25/35/65, 3/10/60/25, 4/45/15/50}{
  \fill[black!\rowa] ({0.28+\col*0.24},0.92) rectangle ({0.52+\col*0.24},1.16);
  \fill[black!\rowb] ({0.28+\col*0.24},0.68) rectangle ({0.52+\col*0.24},0.92);
  \fill[black!\rowc] ({0.28+\col*0.24},0.44) rectangle ({0.52+\col*0.24},0.68);
}
\draw[black!50, line width=0.4pt, step=0.24, shift={(0.28,0.44)}] (0,0) grid (1.2,0.72);
\node[anchor=west, font=\footnotesize] at (1.6,0.98) {$\bm{\rho}\in[0,1]^{m\times n}$};
\node[anchor=west, font=\scriptsize, text=black!55] at (1.6,0.6) {upstream scores};

\node[font=\scriptsize\itshape, text=black!60] at (6.975,5.95) {Co-occur in practice - studied in isolation in prior work};

\draw[rounded corners=3pt, draw=hetA, fill=hetA!5, line width=0.7pt] (4.3,4.15) rectangle (9.65,5.65);
\node[tagA] at (4.65,5.34) {A};
\node[anchor=west, font=\footnotesize\bfseries] at (4.87,5.34) {Allocation structure};
\node[anchor=west, font=\footnotesize] at (4.5,4.84) {$w_{ij}\in\{0,1\}, \quad \textstyle\sum_{j} w_{ij}=k$};
\node[anchor=west, font=\scriptsize, text=black!55] at (4.5,4.44) {playlists, not single picks};
\pic[black!60] at (7.9,4.64) {person};
\draw[arr, line width=0.5pt] (8.12,4.64) -- (8.28,4.64);
\foreach \s in {0,1,2}{
  \draw[draw=hetA, fill=white, line width=0.5pt, rounded corners=1pt] ({8.34+\s*0.38},4.45) rectangle ({8.68+\s*0.38},4.83);
  \pic[hetA, scale=0.62] at ({8.47+\s*0.38},4.58) {note};
}
\node[font=\scriptsize, text=black!55] at (8.9,5.08) {$k$ slots};

\draw[rounded corners=3pt, draw=hetB, fill=hetB!5, line width=0.7pt] (4.3,2.25) rectangle (9.65,3.75);
\node[tagB] at (4.65,3.44) {B};
\node[anchor=west, font=\footnotesize\bfseries] at (4.87,3.44) {Consumer populations};
\node[anchor=west, font=\footnotesize] at (4.5,2.94) {$\mathrm{CVaR}_\alpha$ over losses $\mathcal{L}^{\mathrm{Rel}}_{g}$};
\node[anchor=west, font=\scriptsize, text=black!55] at (4.5,2.54) {some groups are harder to serve};
\draw[black!40, line width=0.4pt] (8.5,2.5) -- (9.5,2.5);
\fill[black!45] (8.6,2.5) rectangle (8.82,3.25);
\fill[black!30] (8.9,2.5) rectangle (9.12,3.05);
\fill[hetC]    (9.2,2.5) rectangle (9.42,2.82);
\draw[arr, draw=hetB, line width=0.7pt] (9.31,2.88) -- (9.31,3.24);

\draw[rounded corners=3pt, draw=hetC, fill=hetC!5, line width=0.7pt] (4.3,0.35) rectangle (9.65,1.85);
\node[tagC] at (4.65,1.54) {C};
\node[anchor=west, font=\footnotesize\bfseries] at (4.87,1.54) {Platform objectives};
\node[anchor=west, font=\footnotesize] at (4.5,1.04) {$\textstyle\sum_{i,j} w_{ij}\rho_{ij}v_j \,\ge\, \theta\, V^{\mathrm{GMV}}_{\max}$};
\node[anchor=west, font=\scriptsize, text=black!55] at (4.5,0.64) {fairness must remain beneficial for business};
\draw[draw=black!50, line width=0.5pt, rounded corners=1pt] (8.4,0.78) rectangle (9.5,1.04);
\fill[hetC!60] (8.42,0.80) rectangle (9.16,1.02);
\draw[dashed, draw=hetC, line width=0.7pt] (9.32,0.66) -- (9.32,1.16);
\node[font=\scriptsize, text=hetC] at (9.32,1.34) {$\theta$};

\draw[rounded corners=3pt, draw=black!75, fill=black!2, line width=1.0pt] (10.3,0.35) rectangle (15.5,5.65);
\node[font=\footnotesize\bfseries] at (12.9,5.32) {This work: study all three jointly};
\node[font=\scriptsize\itshape, text=black!55] at (12.9,4.96) {one optimization lens, same allocation $\bm{w}$};
\draw[black!30] (10.5,4.7) -- (15.3,4.7);

\node[tagB] at (10.68,4.26) {B};
\node[anchor=west, font=\footnotesize] at (10.92,4.26) {$\min_{\bm{w},\,\tau}\ \ \mathrm{CVaR}_\alpha\big(\{\mathcal{L}^{\mathrm{Rel}}_{g}\}_{g=1}^{G}\big)$};
\node[anchor=west, font=\scriptsize\itshape, text=black!55] at (10.68,3.84) {subject to};
\node[tagA] at (10.68,3.44) {A};
\node[anchor=west, font=\footnotesize] at (10.92,3.44) {$w_{ij}\in\{0,1\},\ \ \textstyle\sum_j w_{ij}=k\ \ \forall i$};
\node[rounded corners=1.5pt, draw=black!45, fill=black!8, font=\tiny, text=black!60, inner sep=1.8pt] at (10.68,2.96) {std};
\node[anchor=west, font=\footnotesize] at (10.92,2.96) {$\textstyle\sum_i w_{ij}\ \ge\ \gamma\, U^{*\mathcal{P}}_{\min}\ \ \forall j$};
\node[tagC] at (10.68,2.48) {C};
\node[anchor=west, font=\footnotesize] at (10.92,2.48) {$\textstyle\sum_{i,j} w_{ij}\rho_{ij}v_j\ \ge\ \theta\, V^{\mathrm{GMV}}_{\max}$};
\draw[black!30] (10.5,2.12) -- (15.3,2.12);
\node[anchor=west, font=\scriptsize\bfseries] at (10.5,1.84) {trade-off axes we sweep:};
\node[anchor=west, font=\scriptsize] at (10.5,1.5) {$k$ slots\,\textperiodcentered\,$\alpha$ risk\,\textperiodcentered\,$\gamma$ exposure\,\textperiodcentered\,$\theta$ GMV};
\node[anchor=west, font=\scriptsize, text=black!55] at (10.5,1.12) {instruments, not contributions:};
\node[anchor=west, font=\scriptsize, text=black!55] at (10.5,0.8) {exact MIP\,\textperiodcentered\,LP rounding\,\textperiodcentered\,gradients};

\draw[rounded corners=3pt, draw=black!60, fill=white, line width=0.8pt, densely dashed] (16.1,0.35) rectangle (19.55,5.65);
\node[font=\footnotesize\bfseries] at (17.825,5.32) {Normative goal};
\node[font=\scriptsize\itshape, text=black!55] at (17.825,4.94) {are we there?};
\pic[hetA] at (16.34,4.3) {goalbox};
\node[anchor=west, font=\footnotesize, text=black!75] at (16.58,4.3) {$k$ relevant songs each};
\pic[hetB] at (16.34,3.55) {goalbox};
\node[anchor=west, font=\footnotesize, text=black!75] at (16.58,3.55) {niche groups served};
\pic[black!50] at (16.34,2.8) {goalbox};
\node[anchor=west, font=\footnotesize, text=black!75] at (16.58,2.8) {every artist heard};
\pic[hetC] at (16.34,2.05) {goalbox};
\node[anchor=west, font=\footnotesize, text=black!75] at (16.58,2.05) {platform stays viable};
\node[font=\scriptsize\itshape, text=black!55, align=center] at (17.825,1.05) {not for free:\\fairness costs grow with $k$};

\draw[arr] (3.76,4.9) -- (4.24,4.9);
\draw[arr] (3.76,3.0) -- (4.24,3.0);
\draw[arr] (3.76,1.1) -- (4.24,1.1);
\draw[arr] (9.71,4.9) -- (10.24,3.95);
\draw[arr] (9.71,3.0) -- (10.24,3.0);
\draw[arr] (9.71,1.1) -- (10.24,2.05);
\draw[arr, densely dashed] (15.56,3.0) -- (16.04,3.0);

\end{tikzpicture}%
}
\vspace{-2mm}
\caption{What does fairness cost in a \emph{realistic} two-sided marketplace? Illustrated on music streaming: an upstream recommender supplies relevance scores $\bm{\rho}$, and listeners, artists, and the platform interact through the allocation $\bm{w}$. Real marketplaces exhibit three co-occurring facets of heterogeneity that prior work studies in isolation: \textcolor{hetA}{\textbf{(A)}} discrete multi-item feeds, \textcolor{hetB}{\textbf{(B)}} heterogeneous consumer groups, and \textcolor{hetC}{\textbf{(C)}} business constraints: on top of the standard producer exposure floor ($\gamma$). Rather than proposing a new ranking method, we \emph{study} the three facets jointly through a single optimization lens, sweeping the trade-off axes ($k,\alpha,\gamma,\theta$) with exact and approximate solvers as instruments, and ask how close such a marketplace can come to the normative goal of a feed that serves every side. Our analysis shows it gets there only partly: ``free'' producer fairness at $k{=}1$ disappears once $k>1$.}
\Description{Block diagram. A realistic music-streaming marketplace with heterogeneous listeners (mainstream, casual, niche groups) and artists (popular, long-tail) exhibits three co-occurring facets of heterogeneity, shown as three colored boxes: discrete k-item allocation, consumer-group fairness via CVaR, and a GMV business floor. This work studies all three jointly through one optimization lens, sweeping trade-off axes k, alpha, gamma, theta with exact and approximate solvers as instruments. A dashed box on the right shows the normative goal, a feed fair to all sides, with the question: are we there? Answer: not for free, fairness costs grow with k.}
\label{fig:pipeline}
\vspace{-2mm}
\end{figure*}

 Recent work has recognized this tension and proposed fairness constraints that guarantee minimum exposure to producers \cite{biswasFairRecommendationTwoSided2022, greenwoodUseritemFairnessTradeoffs, 2506.01178}. A striking finding emerged: under certain conditions, producer fairness comes at no cost to consumer utility \citep{greenwoodUseritemFairnessTradeoffs}. This ``free fairness'' result is intuitive in narrow settings. Imagine a platform with 1,000 listeners and only 10 artists, where each listener receives a single song recommendation. Every listener can get their top choice while still ensuring each artist reaches at least 100 listeners, simply because there are far more recommendation slots than artists. However, this favorable regime depends on two assumptions: single-item recommendations ($k=1$) and extreme consumer-to-producer imbalance ($m \gg n$). Real platforms violate both.

 However, the defining characteristic of practical two-sided markets is \emph{heterogeneity} that manifests itself at many levels. There is heterogeneity in incentives: producers want exposure, consumers want relevance. Moreover, real platforms recommend multiple items per consumer: Netflix shows rows of titles, Amazon displays product grids, Spotify generates playlists. This multi-item setting ($k > 1$) creates heterogeneity in allocation complexity. Consumer populations are heterogeneous: users cluster by behavior, demographics, or preference structure, and certain groups may be systematically harder to serve. Producer populations are heterogeneous: items vary in quality, price, and strategic importance to the platform. Business objectives introduce heterogeneity in constraints: platforms must balance consumer satisfaction, producer fairness, and commercial viability simultaneously. \\

\noindent 
Prior work has addressed these dimensions in isolation. Fair allocation approaches in two-sided markets provide theoretical guarantees \cite{2506.01178, biswasFairRecommendationTwoSided2022}, but often do so in constrained or stylized settings, for example via indivisible-good recommendation models or near-feasible allocations obtained by rounding fractional solutions. Intersectional fairness approaches \cite{2402.02816} reveal that aggregate two-sided fairness can mask disparities at attribute intersections, but they are developed for Top-N recommendation rather than tractable optimization of discrete allocation problems. Recent fair re-ranking methods such as CPFair \cite{naghiaei2022cpfair}, P-MMF \cite{xu2023pmmf}, ProFairRec \cite{qi2022profairrec}, and taxation-based re-ranking \cite{xu2024taxation} study provider or multi-stakeholder fairness in realistic ranked recommendation settings, but they typically optimize one fairness notion at a time and do not ask how producer fairness, group-level consumer fairness, and business constraints interact within one shared formulation. However, a unified study to understand these trade-offs remain limited. Our goal in this paper is, therefore, is to study a unified problem framing for evaluating these coupled trade-offs.

 A deeper limitation cuts across much of the literature: consumer objectives are often defined at the individual level, or group fairness is studied separately from producer-side constraints. Whether optimizing mean utility or worst-case individual utility, these formulations can treat consumers as interchangeable. However, a system that achieves high average utility while consistently under-serving users with niche interests is unfair in a way that individual-level objectives cannot detect. Our goal is to provide a faithful and explicit formulation of two-sided markets under these interacting forms of heterogeneity, and to study what this formulation reveals.\\

\noindent \textbf{Contributions.}
 We frame this work as an analytical, foundational study of how producer fairness, consumer-group fairness, and business objectives behave when coupled under a single optimization lens, rather than as a new production-ready ranking architecture. The value of our contribution lies in the joint formulation not in the individual building blocks, which were previously discussed. Concretely, we establish the following: 

\begin{itemize}[leftmargin=15pt, topsep=0.4em, itemsep=0.4em]

    \item  \textbf{Free fairness breaks down in multi-item settings.} Across three datasets, the ``free fairness'' result \citep{greenwoodUseritemFairnessTradeoffs} does not generalize: enforcing producer fairness is essentially costless at $k=1$ but incurs a growing consumer-utility cost as $k$ increases, with drops of 15-25\% at $k=10$ under full fairness.
    
    \item  \textbf{A unified formulation of two-sided heterogeneity.} We formalize two-sided fairness under three sources of heterogeneity usually studied separately—discrete multi-item allocation ($w_{ij}\in\{0,1\}$, $k>1$), consumer-group fairness via Conditional Value at Risk (CVaR), and business (GMV) constraints—within one optimization view. Even though each ingredient is individually known, making their interaction explicit and jointly analyzable is what lets us observe trade-offs that component-wise studies cannot.
    
    \item  \textbf{Empirical characterization of the coupled trade-offs.} Using exact mixed-integer programming (MIP), LP relaxation with rounding, and gradient-based solvers as instruments—not as new ranking algorithms—we show that CVaR compresses group-level disparities and that moderate producer-fairness constraints can even improve simulated business metrics by diversifying exposure away from saturated producers. We present these as evidence about the trade-offs induced by the formulation, not as a claim of state-of-the-art superiority over the broader fair re-ranking literature.
    
\end{itemize}

\noindent  The remainder of the paper is organized as follows. Section~2 reviews related work. Section~3 formalizes the problem, identifying where prior formulations fall short and introducing our extensions aware of heterogeneity. Section~4 develops scalable optimization methods. Section~5 presents experiments, and Section~6 concludes.

\section{Related Work}

 Fairness-aware recommendation has expanded from single-stakeholder utility optimization to multi-stakeholder settings in which consumers, producers, and platforms have competing objectives \cite{li2023fairness, deldjoo2024fairness, wang2023survey, abdollahpouri2020multistakeholder}. In two-sided marketplaces, this tension is commonly formalized as balancing consumer relevance against provider exposure or opportunity \cite{zheng2017multi, singh2018fairness, greenwoodUseritemFairnessTradeoffs}.

 Previous literature has aimed to quantify such tension in a controlled setting. For example, Greenwood et al. \cite{greenwoodUseritemFairnessTradeoffs} show that provider fairness can be ``free'' under a stylized $k=1$ regime, while subsequent work extends provider-aware re-ranking in practical recommendation settings, for example through personalized consumer/producer trade-offs \cite{naghiaei2022cpfair}, provider max-min fairness \cite{xu2023pmmf}, fairness-aware news recommendation \cite{qi2022profairrec}, taxation-inspired re-ranking \cite{xu2024taxation}, and broader toolkits for fairness and diversity evaluation \cite{xu2025fairdiverse}. These studies establish strong baselines and design patterns for provider-aware ranking. Our paper builds on this literature, but focuses on a different question: how do producer fairness, consumer-group fairness, and business constraints interact when treated as parts of a single optimization problem?
 
On the relevant question of  \textit{group fairness and risk-sensitive objectives on the consumer side}, subgroup and intersectional fairness formulations show that average performance can mask systematic harms to disadvantaged groups \cite{2106.02702, 2402.02816}. CVaR and related risk measures are well-established tools for protecting the lower tail of an outcome distribution, and have been used in fairness-aware machine learning more broadly \cite{williamsonFairnessRiskMeasures2019}.However, the study of how such CVaR risk measure interacts within the richer two-sided markets setting is limited. Our contribution is to use it as the consumer-side component of a two-sided recommendation formulation that also includes provider guarantees and business constraints, making those trade-offs explicit and measurable in one framework.

 Additionally, re-ranking methods have long incorporated platform objectives such as revenue, diversity, or strategic value alongside relevance \cite{ribeiroMultiobjectiveParetoEfficientApproaches2015, nguyenMultiObjectiveLearningReRank2017, basuFrameworkFairnessTwoSided2020}. Related assortment and visibility constrained optimization work similarly emphasizes that platform design rarely reduces to a single objective \cite{2307.13656}. Our GMV-constrained formulation belongs to this tradition. The difference is that we integrate the business term into the same framework used for producer and group-level consumer fairness, which lets us characterize a three-way trade-off rather than evaluating metrics purely post hoc.

 Taken together, these literature motivate our positioning. We do not argue that fairness-aware re-ranking, provider fairness, or CVaR-based group protection are individually new. Instead, we position the paper as a unifying formulation and empirical study of what changes when these ingredients are combined under realistic heterogeneity: discrete multi-item allocation, producer-side guarantees, consumer-group protection, and business-aware constraints. This framing also clarifies the paper's scope: it is an offline optimization framework and analysis tool for studying trade-offs, not a claim that one global solver is the final deployment recipe for all large-scale recommender systems.

\section{Problem Formulation}

 This section formalizes the two-sided recommendation problem. We begin with notation and the standard formulation inherited from prior work, then identify where this formulation falls short in capturing the heterogeneity of practical platforms. Finally, we formalize our proposal for a unified treatment of the different facets of heterogeneity in two-sided markets: discrete multi-item allocations, group-level consumer fairness, and business constraints.

\subsection{The Standard Two-Sided Formulation}

 We consider the final allocation stage of a recommendation pipeline, operating as a fairness-aware \emph{re-ranking} layer. An upstream retrieval model has produced a relevance score matrix $\pmb{\rho} = (\rho_{ij}) \in [0,1]^{m \times n}$ between $m$ consumers and $n$ producers \cite{peng2022survey, jadon2024comprehensive}. The task is to construct an allocation matrix $\pmb{w} = (w_{ij})$ that assigns producers to consumers while balancing competing objectives.

 Following prior work on two-sided fairness \cite{greenwoodUseritemFairnessTradeoffs, biswasFairRecommendationTwoSided2022}, we adopt simplifying assumptions that isolate the allocation problem from confounding factors:

\begin{itemize}[leftmargin=15pt, topsep=0.4em, itemsep=0.4em]

\item[\textbf{(A1)}]  \textbf{No positional bias.} All slots in a recommendation list receive equal attention. This assumption isolates the allocation decision from presentation effects, allowing us to study fairness in item selection without conflating it with ranking optimization. While real systems exhibit position bias, addressing it requires jointly modeling attention and allocation, which we leave to future work.

\item[\textbf{(A2)}]  \textbf{Relevance determines transactions.} The probability that consumer $i$ engages with producer $j$ is proportional to their relevance score $\rho_{ij}$. This connects the allocation problem to downstream outcomes: maximizing relevance correlates with maximizing engagement and transactions.

\item[\textbf{(A3)}]  \textbf{One item per producer.} Each producer offers a single item. This avoids intra-producer competition and simplifies the fairness constraint to producer-level exposure rather than item-level exposure.

\item[\textbf{(A4)}]  \textbf{Unit supply.} Each item can be recommended to multiple consumers without depletion during the allocation phase. This is appropriate for digital goods (movies, songs, articles) and for physical goods where inventory constraints are handled separately from recommendation.

\end{itemize}

\noindent  These assumptions are deliberate scoping choices that let us isolate the core tension between consumer and producer objectives while keeping the framework analytically tractable. They do not aim to fully capture deployment settings; rather, they define the baseline formulation against which we later introduce additional heterogeneity through multi-item allocation, group-level fairness, and business constraints. Two-sided markets have two primary stakeholders with conflicting incentives, and we define a utility function for each. \\

\noindent \textbf{Consumer Utility.}  A consumer's utility is the relevance captured by their recommended set, normalized by the consumer's single best achievable relevance. For consumer $i$ receiving allocation $\pmb{w}$:
\begin{equation}
U^{\mathcal{C}}_{i}(\pmb{\rho}, \pmb{w}) = \frac{\sum_{j} w_{ij}\rho_{ij}}{\max_j \rho_{ij}},
\label{eq:consumer-util}
\end{equation}
which matches the objective used in our exact max-min and mean formulations, and preserves comparability with prior single-item work \cite{greenwoodUseritemFairnessTradeoffs} even when we move to multi-item allocations. Under this normalization, utility can exceed $1.0$ when $k > 1$, so it should be interpreted as relevance accumulated relative to the consumer's best single-item match rather than as a fraction of the top-$k$ optimum. For the CVaR formulation introduced below, however, we switch to a top-$k$ consumer-greedy baseline so that group losses are measured relative to the best achievable $k$-item relevance for each consumer. \\

\noindent \textbf{Producer Utility.} A producer's utility is total exposure across consumers, denoted
\begin{equation}
U^{\mathcal{P}}_{j}(\pmb{w}) = \sum_{i} w_{ij}.
\label{eq:producer-util}
\end{equation}
This captures the producer's incentive for visibility but may conflict with consumer relevance. \\

\noindent \textbf{Balancing Competing Objectives.} Since consumer and producer utilities are inherently in tension, we need a mechanism to balance them. A standard approach is constrained optimization with minimum exposure guarantees. The two-sided fairness framework of \cite{greenwoodUseritemFairnessTradeoffs} maximizes minimum consumer utility subject to a producer exposure constraint:
\begin{align}
\max_{\pmb{w} \in [0,1]^{m \times n}} \; & \min_i \sum_j \frac{w_{ij}\rho_{ij}}{\max_j \rho_{ij}} \nonumber \\
\text{s.t.} \quad & \min_j \sum_i w_{ij} \ge \gamma \cdot U^{*\mathcal{P}}_{\min}, \quad \sum_j w_{ij} \le 1,
\label{eq:greenwood}
\end{align}
where $\gamma \in [0,1]$ modulates producer fairness and $U^{*\mathcal{P}}_{\min}$ is the maximum achievable minimum exposure. This formulation is representative of a general class of methods that use worst-case or minimum-exposure constraints to guarantee producer fairness.

\subsection{Where the Standard Formulation Falls Short}

The formulation in ~\autoref{eq:greenwood} captures the essential consumer-producer tension but makes assumptions that do not hold in practical recommendation systems. We identify three dimensions of heterogeneity that it ignores, elaborated on next. \\

\noindent \textbf{Heterogeneity in Allocation Structure.}
Real platforms recommend \emph{multiple} items per consumer ($k > 1$): Netflix shows rows of titles, Amazon displays product grids, Spotify generates playlists. The standard formulation uses soft allocations ($w_{ij} \in [0,1]$) with at most one item per consumer ($\sum_j w_{ij} \le 1$). Soft allocations are natural in domains like ride-hailing, where a driver can be fractionally assigned across riders (e.g., 40\% probability of serving rider A, 60\% for rider B). In contrast, recommendation is inherently discrete: a movie either appears in a user's top-10 list or it does not. Furthermore, when $k > 1$, each consumer's recommendation list must include items beyond the single best match, creating coupling across consumer choices that does not exist at $k=1$. \\

\noindent \textbf{Heterogeneity in Consumer Populations.}
Standard objectives (mean utility, max-min individual utility) treat consumers as interchangeable. In practice, consumer populations are heterogeneous: users cluster by behavior, demographics, or preference structure. A system that achieves high average utility while consistently under-serving users with niche interests or sparse interaction histories is unfair in a way that individual-level objectives cannot detect. Certain consumer \emph{groups} may be systematically harder to serve, and protecting them requires group-level objectives. \\

\noindent \textbf{Heterogeneity in Platform Objectives.}
From a platform's perspective, fairness is not an end in itself but must coexist with commercial viability. Business objectives such as revenue, transaction value, or sell-through rate introduce additional constraints that the standard formulation ignores. Any fairness intervention that does not account for business incentives is unlikely to be adopted in practice.

\subsection{Heterogeneity-Aware Two Sided Markets}

 We now present our formalization of two-sided markets that remain faithful to the multiple facets of heterogeneity we argued in the previous section. Together, they define a more realistic problem setting for two-sided fairness. \\

\noindent \textbf{Soft Allocations to Discrete Multi-Item Allocations.}
 We replace soft allocations with binary decisions $w_{ij} \in \{0,1\}$ and generalize from single-item to multi-item recommendations, that is
\begin{equation}
\sum_{j} w_{ij} = k, \quad \forall i \in [m].
\label{eq:k-constraint}
\end{equation}
We redefine the producer fairness baseline $U^{*\mathcal{P}}_{\min}$ accordingly, hence
\begin{equation}
U^{*\mathcal{P}}_{\min} = \max_{\pmb{w} \in \{0,1\}^{m \times n}} \min_j \sum_i w_{ij} \quad \text{s.t.} \quad \sum_j w_{ij} = k \; \forall i.
\label{eq:producer-baseline}
\end{equation}
This formulation is a mixed-integer program (MIP), which is NP-hard \cite{korte2011combinatorial}. The computational difficulty reflects genuine problem structure: at $k > 1$, giving producer $j$ to consumer $i_1$ may preclude giving it to consumer $i_2$ if fairness constraints bind, creating complex dependencies that simple heuristics cannot resolve optimally. \\

\noindent \textbf{Aggregate Performance to Group-Level Consumer Fairness via CVaR.}
To protect systematically disadvantaged consumer groups, we target the tail of the group utility distribution using Conditional Value at Risk (CVaR) \cite{rockafellar2000optimization, artzner1999coherent}. CVaR has been proposed as a fairness measure in classification \cite{williamsonFairnessRiskMeasures2019} but has not been applied to two-sided recommendation allocation.

Suppose consumers are partitioned into groups $g_1, \ldots, g_G$, e.g., by behavioral clustering or demographic attributes. For CVaR, we first compute each consumer's greedy $k$-item baseline $g_i = \sum_{\hat{j} \in \text{Top}_k(i)} \rho_{i\hat{j}}$, then define the relevance loss for group $g$ as
\begin{equation}
\mathcal{L}^{\text{Rel}}_g(\pmb{\rho}, \pmb{w}) = \frac{1}{|g|} \sum_{i \in g} \left(1 - \frac{\sum_j w_{ij}\rho_{ij}}{g_i}\right).
\label{eq:group-loss}
\end{equation}
The CVaR objective minimizes expected loss in the worst-$\alpha$ fraction of groups:
\begin{equation}
\min_{\pmb{w}, \tau \ge 0} \; \tau + \frac{1}{(1-\alpha)G} \sum_{g=1}^G \max\{\mathcal{L}^{\text{Rel}}_g - \tau, 0\},
\label{eq:cvar-obj}
\end{equation}
subject to allocation and producer fairness constraints. The parameter $\alpha$ controls risk aversion: $\alpha = 0$ recovers mean optimization, while $\alpha \to 1$ approaches max-min over groups. \\

\noindent \textbf{Normative Requirements to Business Constraints.}
We incorporate business objectives directly into the optimization. Let $v_j$ denote the value of producer $j$ (e.g., price, margin, or strategic importance). We impose a minimum Gross Merchandise Value (GMV) constraint:
\begin{equation}
\sum_{i=1}^{m} \sum_{j=1}^{n} w_{ij} \rho_{ij} v_j \ge \theta \cdot V^{\text{GMV}}_{\max},
\label{eq:gmv-constraint}
\end{equation}
where $\theta \in [0,1]$ is a threshold and $V^{\text{GMV}}_{\max}$ is the maximum achievable GMV. This constraint aligns fairness interventions with platform incentives, allowing navigation of a three-way trade-off between consumer utility, producer fairness, and revenue.

\subsection{Why This Matters: Free Fairness Revisited}

 Having established our heterogeneity-aware formulation, we revisit the ``free fairness'' result. Prior work \cite{greenwoodUseritemFairnessTradeoffs} showed that under their formulation, enforcing producer fairness imposes no cost on consumer utility when $m \gg n$ and $k=1$. The intuition is simple: when each consumer needs only one item and there are far more consumers than producers, there is enough ``slack'' to distribute exposure evenly without sacrificing relevance.

 Our formulation allows us to test whether this result generalizes. As we demonstrate empirically in Section~\ref{sec:experiments} (Figure~\ref{fig:res_optim}), free fairness \emph{does not hold} in realistic multi-item settings. At $k=1$, fairness is indeed nearly free. At $k=5$ or $k=10$, enforcing producer exposure guarantees reduces mean consumer utility by 10-30\% depending on the dataset. The favorable regime that prior work identified is a special case, not the norm. \\

 \noindent \textbf{When should we expect free fairness to hold or fail?} While our evidence is empirical, the mechanism admits an intuitive structural explanation rooted in the relevance matrix $\pmb{\rho}$. At $k=1$, each consumer occupies a single slot, so satisfying a producer's minimum exposure only requires finding \emph{some} set of consumers for whom that producer is a near-top match. When preferences are diverse—$\pmb{\rho}$ is effectively high-rank and consumers spread their top choices across many producers—such consumers almost always exist, and exposure can be redistributed at negligible relevance cost. Free fairness is therefore best understood as a property of \emph{preference diversity} rather than of fairness constraints per se. The argument weakens precisely when this diversity is absent: if $\pmb{\rho}$ is low-rank, or if consumer groups cluster around the same small set of dominant producers, then the consumers who must absorb a constrained producer genuinely prefer something else, and enforcing exposure forces real utility loss. The multi-item regime ($k>1$) amplifies this effect: beyond each consumer's single best match, additional slots must be filled with progressively less relevant items, so the relevance cost of redistributing exposure accumulates across all $k$ slots. This framing suggests the breakdown is intrinsic to multi-item allocation under correlated preferences rather than a dataset artifact, though we leave a formal characterization to future work.

 Furthermore, Figure~\ref{fig:group-fairness} shows that individual-level objectives leave substantial variance across consumer groups. Optimizing for mean or max-min individual utility does not protect systematically disadvantaged groups; CVaR optimization is necessary to compress group-level disparities. These findings underscore our  formalization and the crucial study of two-sided under realistic heterogeneity. 
\section{Methods}

 In the last section, we formalize two-sided fairness under the realistic facets of heterogeneity: discrete multi-item allocations, group-level consumer fairness via CVaR, and business constraints. While our findings reveal the utility of studying two-sided markets under this setting, binary allocations turn fairness-aware recommendation into a mixed-integer program, which is NP-hard. At production scale, exact solvers become impractical, and we need tractable alternatives.

 In this section, we develop practical optimization methods for heterogeneity-aware re-ranking. We first present exact formulations that serve as baselines, then introduce more tractable approximations: LP relaxation with rounding, and differentiable gradient-based optimization. Although each ingredient of the formulation producer fairness constraints, CVaR, and GMV thresholds—has appeared in prior work, our focus here is on instantiating them \emph{jointly} so their interaction can be studied under a common re-ranking layer. We do not claim these methods are immediately production-ready; rather, at the problem sizes we test, the approximations scale better than the exact MIP solver while preserving most of the fairness behavior established in our formulation.

\subsection{Exact Formulations}

 We present two complementary objectives for the discrete multi-item allocation problem. Both extend prior work to the binary setting with $k$ recommendations per consumer. \\

\noindent \textbf{Max-Min Consumer Utility.}
 The first formulation directly generalizes the max-min objective to discrete multi-item allocations:
\begin{align}
\max_{\pmb{w} \in \{0,1\}^{m \times n}} & \min_{i} \frac{\sum_{j=1}^n w_{ij}\rho_{ij}}{\max_{j} \rho_{ij}} \nonumber \\
\text{s.t.} \quad & \sum_{j=1}^n w_{ij} = k, \quad \forall i \in [m], \nonumber \\
& \min_{j} \sum_{i=1}^m w_{ij} \ge \gamma \cdot U^{*\mathcal{P}}_{\min}. \label{eq:maxmin_multi}
\end{align}

\noindent  This guarantees a minimum relevance level for all consumers but is sensitive to outliers and computationally demanding. \\

\noindent \textbf{Mean Consumer Utility.}
 An alternative maximizes mean utility while enforcing the same producer fairness constraint:
\begin{align}
\max_{\pmb{w} \in \{0,1\}^{m \times n}} & \frac{1}{m}\sum_{i=1}^m \frac{\sum_{j=1}^n w_{ij}\rho_{ij}}{\max_{j} \rho_{ij}} \nonumber \\
\text{s.t.} \quad & \sum_{j=1}^n w_{ij} = k, \quad \forall i \in [m], \nonumber \\
& \min_{j} \sum_{i=1}^m w_{ij} \ge \gamma \cdot U^{*\mathcal{P}}_{\min}. \label{eq:avg_multi}
\end{align}

\noindent Together, ~\autoref{eq:maxmin_multi} and \autoref{eq:avg_multi} provide two ends of a spectrum: the former emphasizes worst-case guarantees, while the latter optimizes aggregate satisfaction.

\begin{figure*}[t]
\centering
\includegraphics[width=0.82 \linewidth]{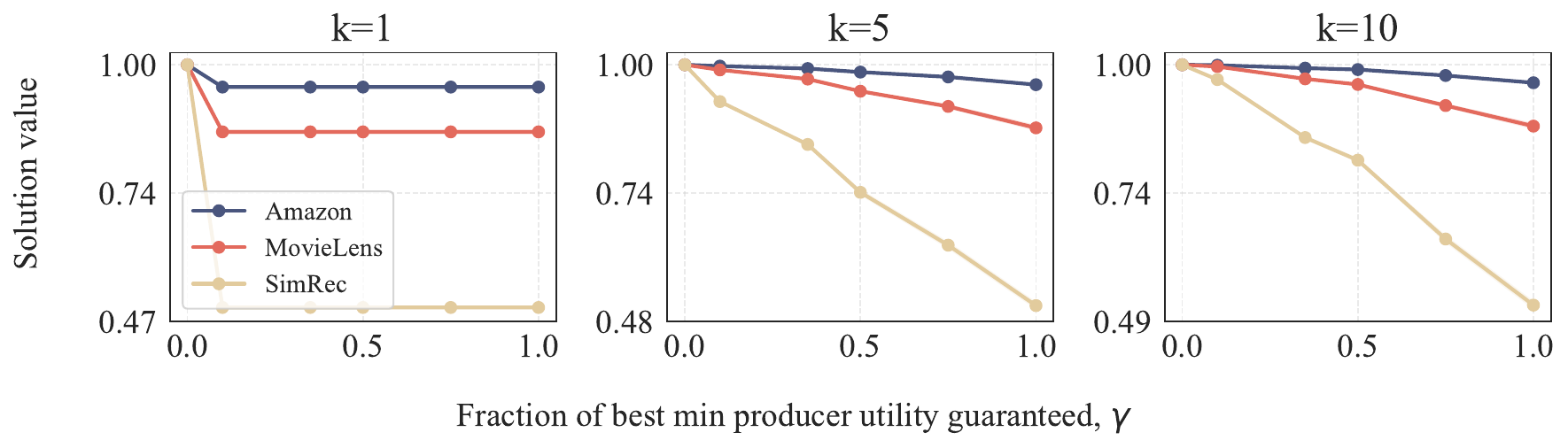}
\vspace{-2mm}
\caption{Mean consumer utility versus producer fairness ($\gamma$) for $k \in \{1, 5, 10\}$ across Amazon Reviews, MovieLens, and SimRec. At $k=1$, utility remains nearly flat as $\gamma$ increases (free fairness). At $k>1$, utility declines with stricter fairness, and the slope steepens with larger $k$.}
\label{fig:res_optim}
\end{figure*}

\subsection{Group Fairness via CVaR}

 Individual-level objectives cannot detect systematic disparities across consumer groups. To capture heterogeneity in consumer populations, we extend the framework to group-level fairness.

 Suppose consumers are partitioned into groups $g_1,\dots,g_G$. Let $g_i = \sum_{\hat{j} \in \text{Top}_k(i)} \rho_{i\hat{j}}$ denote consumer $i$'s greedy top-$k$ baseline. For each group $g$, the relevance loss is:
\begin{equation}
\mathcal{L}^{\text{Rel}}_{g}(\pmb{\rho}, \pmb{w}) = \frac{1}{N_{g}}\sum_{i\in g} \Biggl( 1 - \frac{\sum_j w_{ij}\rho_{ij}}{g_i}\Biggr),
\label{eq:group-loss-method}
\end{equation}
where $N_g$ is the group size and $g_i$ matches the consumer-greedy baseline used in the implementation. We then optimize a CVaR objective over groups:
\begin{equation}
\min_{\pmb{w},\, \tau \ge 0} \quad
\tau + \frac{1}{(1 - \alpha) G} \sum_{g=1}^G \max\{\mathcal{L}^{\text{Rel}}_{g}(\pmb{\rho},\pmb{w}) - \tau,\,0\},
\label{eq:cvar}
\end{equation}
subject to $\sum_j w_{ij} = k$ for all $i$ and producer fairness constraints. The parameter $\alpha$ controls risk aversion, with higher values focusing optimization on the worst-off groups. This shifts the focus from protecting only individual users to ensuring equitable treatment of consumer subgroups, a crucial property when serving heterogeneous populations.

\subsection{Business Constraints: GMV Integration}

 Heterogeneity in platform objectives means fairness must coexist with commercial viability. We integrate Gross Merchandise Value (GMV) as a constraint, inspired by multi-objective re-ranking methods \cite{nguyenMultiObjectiveLearningReRank2017}. Each producer $j$ has value $v_j$ (e.g., price or margin). The maximum achievable GMV is:
\begin{equation}
V_{\max}^{\mathrm{GMV}} = \max_{\pmb{w} \in \{0,1\}^{m \times n}} \sum_{i=1}^{m} \sum_{j=1}^{n} w_{ij} \rho_{ij} v_j
\quad \text{s.t.} \quad \sum_{j=1}^{n} w_{ij} = k, \; \forall i \in [m],
\label{eq:gmv-max}
\end{equation}
which is the highest GMV attainable under the same cardinality constraints. We require that the allocation achieves at least a fraction $\theta \in [0,1]$ of this maximum:
\begin{equation}
\sum_{i=1}^{m} \sum_{j=1}^{n} w_{ij} \rho_{ij} v_j \ge \theta \cdot V_{\max}^{\mathrm{GMV}},
\label{eq:gmv}
\end{equation}
while maintaining the $k$-allocation constraint. The GMV constraint can be combined with any fairness objective (max-min, mean, or CVaR), allowing platforms to navigate the three-way trade-off between consumer utility, producer fairness, and revenue.

\begin{figure}[t]
\centering
\includegraphics[width=0.75\linewidth]{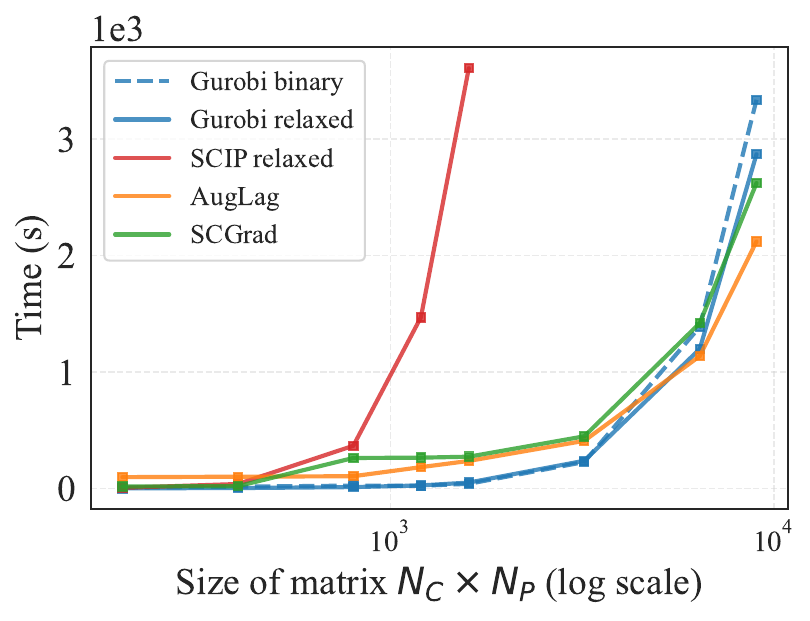}
\vspace{-2mm}
\caption{Runtime comparison of various optimization methods on the Amazon Reviews dataset under $\gamma = 0.5$, $\alpha = 0.95$, and $k = 10$ as a function of increasing relevance matrix size.}
\label{fig:times}
\end{figure}

\subsection{Scalable Optimization Methods}

 Binary allocations make fairness-aware recommendation NP-hard. While solvers like Gurobi \cite{gurobi} can handle moderate problem sizes, large-scale re-ranking requires approximate methods. To this end, we investigate two strategies. \\

\noindent \textbf{Linear Program Relaxation with Rounding.}
 We relax binary decisions $w_{ij}\in\{0,1\}$ to $w_{ij}\in[0,1]$, yielding a linear program (LP). After solving, we recover binary allocations $\hat{\pmb{w}}$ through rounding. We evaluate three schemes:

\begin{itemize}[leftmargin=15pt, topsep=0.4em, itemsep=0.4em]

    \item \textbf{Hard thresholding:} $\hat{w}_{ij}=1$ if $w_{ij}\geq 0.5$, else 0.
    
    \item \textbf{Probabilistic rounding:} $\hat{w}_{ij}\sim \mathrm{Bernoulli}(w_{ij})$, averaged over samples.
    
    \item \textbf{Top-$k$ selection:} Assign the $k$ largest weights per consumer.
    
\end{itemize}

\noindent  LP relaxations provide fast approximations, but rounding may degrade fairness guarantees. \\

\noindent \textbf{Gradient-Based Optimization.}
 To exploit GPU acceleration, we directly optimize continuous allocations parameterized as $w_{ij} = \sigma(z_{ij}/\eta^{(t)})$, with temperature $\eta^{(t)}$ annealed to encourage near-binary outputs. Our main objective is the CVaR loss given by
\begin{equation}
\mathcal{L}_{\mathrm{util}}(\pmb{\rho},\pmb{w},\tau) =
\tau + \tfrac{1}{(1-\alpha)G}\sum_{g=1}^G
\max\{\mathcal{L}^{\text{Rel}}_{g}(\pmb{\rho},\pmb{w})-\tau,0\},
\label{eq:cvar-loss}
\end{equation}
and we explore two optimization strategies:

\begin{itemize}[leftmargin=15pt, topsep=0.4em, itemsep=0.4em]

    \item \textbf{Augmented Lagrangian (AugLag):} Adds quadratic penalties and dual variables to enforce constraints. Provides stricter feasibility but requires careful tuning.
    
    \item \textbf{Soft-Constrained Gradient (SCGrad):} Treats constraints as soft penalties without dual variables. Lighter-weight and converges smoothly, offering a practical trade-off between guarantees and scalability.
    
\end{itemize}

\noindent  We provide detail formulations in the Appendix \ref{sec:optimization-frameworks}. Together, at the scales we evaluate, these methods make heterogeneity-aware re-ranking more tractable than exact MIP while preserving most of the core fairness properties established in our formulation; we do not claim feasibility under industrial-scale latency requirements.

\section{Experimental Validation}
\label{sec:experiments}

 We now empirically validate our heterogeneity-aware formulation across three datasets. Our experiments aim to address four key questions: \emph{(1)} Does free fairness hold in multi-item settings? \emph{(2)} Does CVaR reduce group-level disparities? \emph{(3)} Can fairness improve business metrics? And finally \emph{(4)} do scalable methods match exact solvers? The answers reveal that heterogeneity fundamentally changes the fairness-utility landscape and that our formulation captures trade-offs that prior work missed. \\

\noindent \textbf{Experimental Design.}  We evaluate on two real-world benchmarks, MovieLens-100k \cite{harper2015movielens} and Amazon Reviews \cite{hou2024bridging}, and a synthetic dataset SimRec designed to probe fairness under controlled disparities. For the real datasets, we train a two-tower recommender \cite{10.1145/2959100.2959190} on user-item interactions and compute full consumer-producer relevance matrices $\pmb{\rho}$. This yields $\pmb{\rho}\in[0,1]^{943\times 1680}$ for MovieLens and $\pmb{\rho}\in[0,1]^{10000\times 10000}$ for Amazon Reviews. The recommender achieves high top-$k$ quality (see Table \ref{tab:rec-res-appendix}), ensuring downstream allocation experiments operate on meaningful relevance scores.

 SimRec simulates $1000$ consumers and $1000$ producers with group heterogeneity. Consumers are partitioned into $G=10$ groups drawn from a Zipf distribution, mimicking imbalanced populations. Each consumer's relevance curve follows a smoothed log-decay with additive Gaussian noise, producing a structured $\pmb{\rho}\in[0,1]^{1000\times 1000}$ with ground-truth group structure. Users are clustered into groups reflecting heterogeneity: genre-based clusters for MovieLens (9 groups), category-based clusters for Amazon (18 groups), and predefined groups in SimRec. We solve discrete allocation problems with SCIP \cite{bolusani2024scip} and Gurobi \cite{gurobi}, and run relaxed and gradient-based methods on CPUs and GPUs. Experiments are parallelized over up to 128 cores and A100 GPUs to ensure comparability.

 All code, datasets, and experimental configurations are available at \url{https://github.com/dqmis/equity-by-design}. The repository includes implementations of all optimization methods, data preprocessing pipelines, and scripts to reproduce figures and tables.

\begin{figure}[t]
\centering
\includegraphics[width=1.1\linewidth]{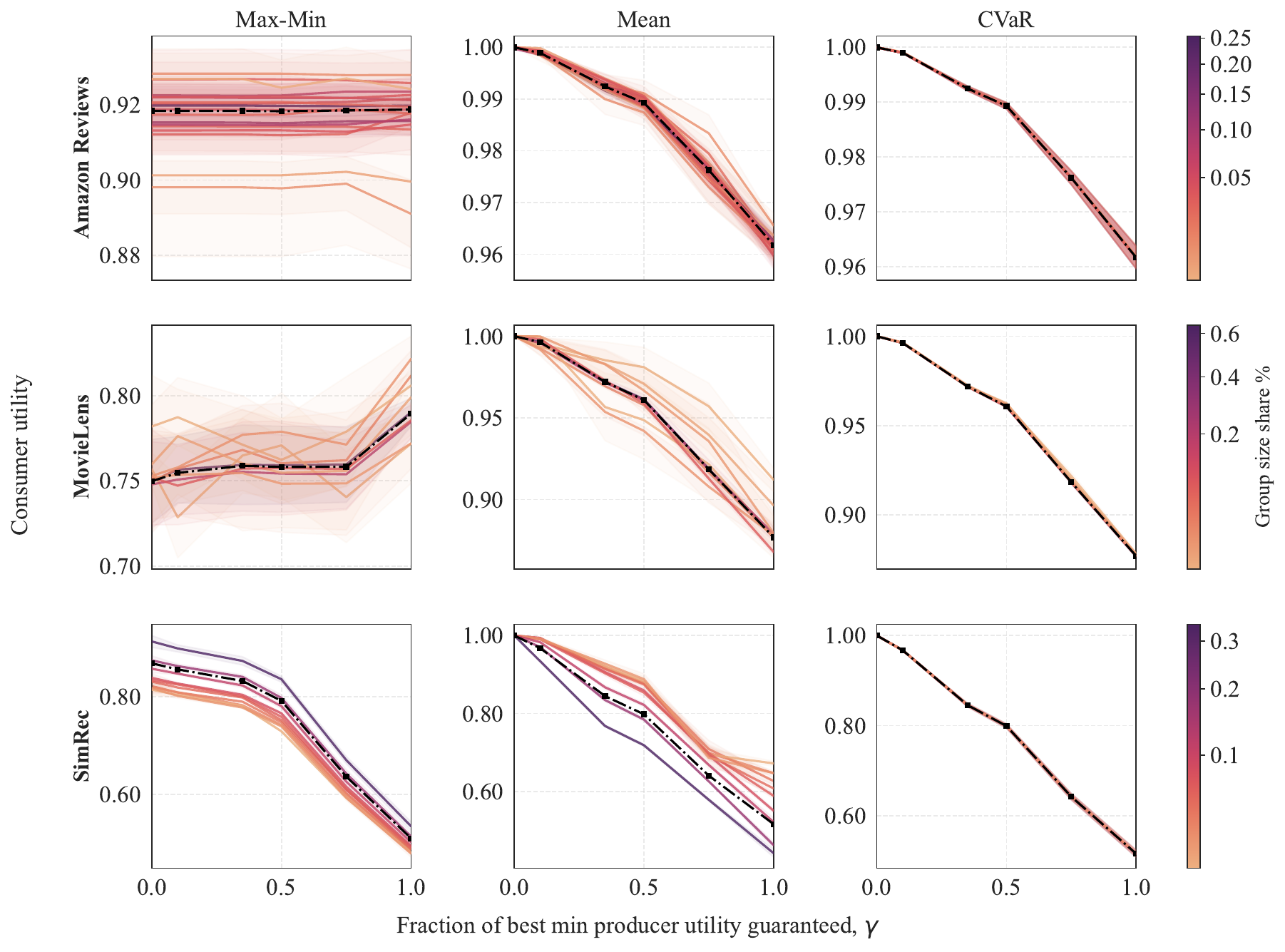}
\caption{Group-level consumer utilities under max-min, mean, and CVaR optimization across $\gamma$. Lines represent consumer groups, colored by share of total sample (darker = larger).}
\label{fig:group-fairness}
\end{figure}

\subsection{Evaluation Results}

\noindent \textbf{Fairness-Utility Trade-offs.} A central question in two-sided fairness is whether producer exposure guarantees must come at the expense of consumer utility. Prior work \cite{greenwoodUseritemFairnessTradeoffs} established a surprising ``free fairness'' result: in single-item settings ($k=1$) with imbalanced consumer-producer ratios, enforcing minimum producer exposure imposes no cost on consumer utility. As stated before, the intuition is that when each consumer needs only one item and consumers vastly outnumber producers, the system has enough flexibility to distribute exposure evenly while still matching each consumer to a highly relevant item.

 We test whether this result generalizes to multi-item recommendations. Figure~\ref{fig:res_optim} shows mean consumer utility as a function of the producer fairness parameter $\gamma$ (where $\gamma=0$ ignores producer fairness and $\gamma=1$ enforces maximum equitable exposure) across three datasets and recommendation list lengths $k \in \{1, 5, 10\}$. \\

\noindent  The results reveal a clear pattern. At $k=1$, the curves are nearly flat across all datasets: consumer utility remains high regardless of $\gamma$, confirming the free fairness phenomenon. However, as $k$ increases to 5 and 10, the curves develop a pronounced downward slope. Enforcing producer fairness now requires recommending items beyond each consumer's top choices, and the relevance cost accumulates across all $k$ slots. At $k=10$ with $\gamma=1$, mean utility drops by 15--25\% relative to the unconstrained baseline ($\gamma=0$), depending on the dataset.

\begin{figure}[t]
\centering
\includegraphics[width=.94\linewidth]{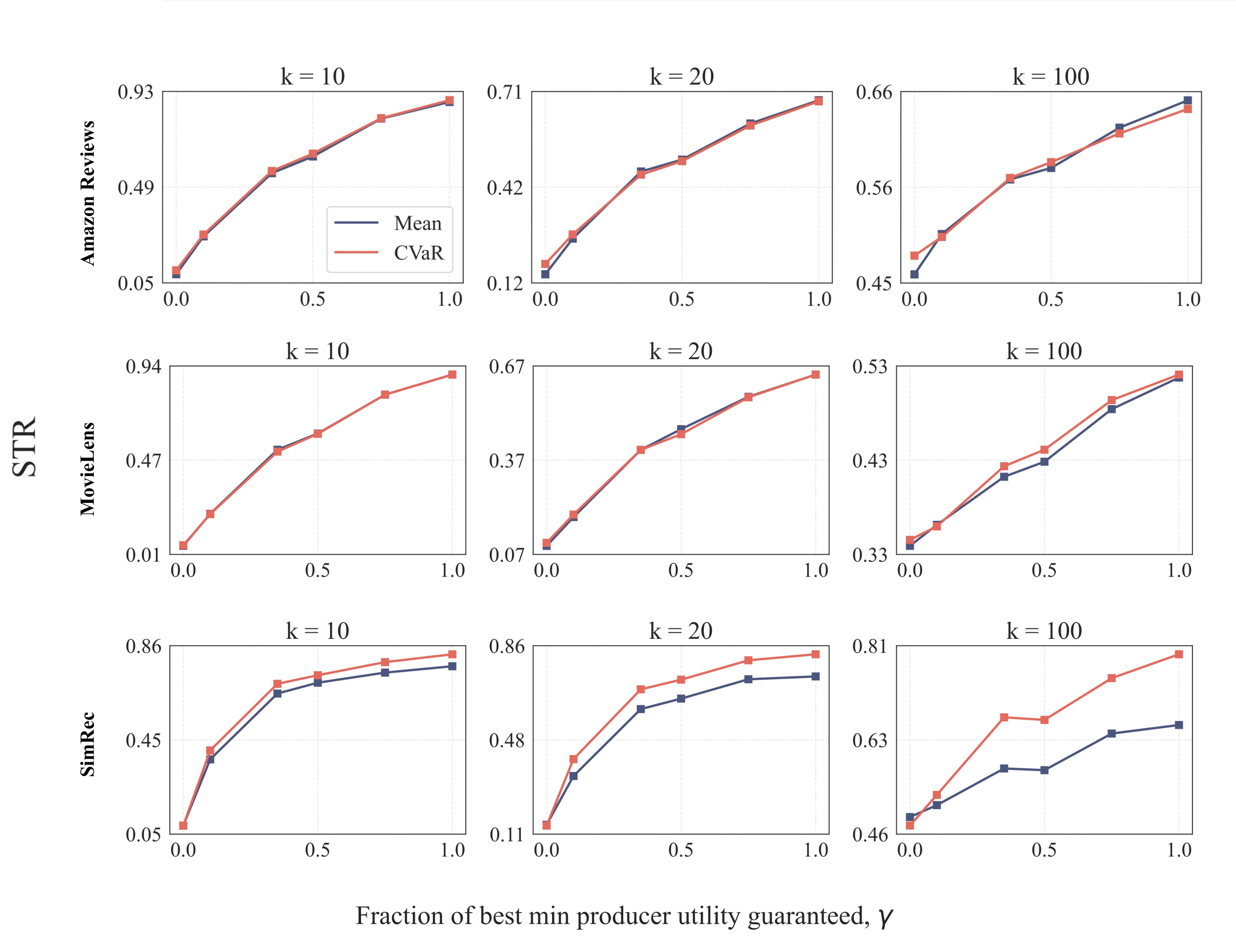}
\caption{Simulated sell-through rate (STR) as a function of producer fairness level $\gamma$ using mean and CVaR consumer utility objectives.}
\label{fig:sim-str}
\end{figure}

 We argue this finding has practical implications. Real recommendation systems rarely show a single item; they present lists, grids, or feeds containing tens of items. In these settings, the free fairness result does not apply. Platform designers must explicitly navigate the trade-off between consumer relevance and producer exposure, and the cost of fairness grows with recommendation numbers per user. \\

\noindent \textbf{Group Fairness with CVaR.}  We next evaluate Conditional Value at Risk (CVaR) optimization, which explicitly controls group-level risk. Figure~\ref{fig:group-fairness} confirms that CVaR consistently reduces inter-group variance compared to both mean and max-min optimization. In SimRec, CVaR balances disadvantaged and advantaged groups, while in MovieLens higher producer fairness $\gamma$ amplifies this effect by indirectly improving overall consumer equity. \\

\noindent \textbf{Fairness and Business Alignment.}  Finally, we test whether fairness can enhance business outcomes. We simulate purchases under supply constraints, where consumers probabilistically buy from their recommended $k$ items, and evaluate Sell-Through Rate (STR) and Gross Merchandise Value (GMV).

\begin{figure}[t]
\centering
\includegraphics[width=1\linewidth]{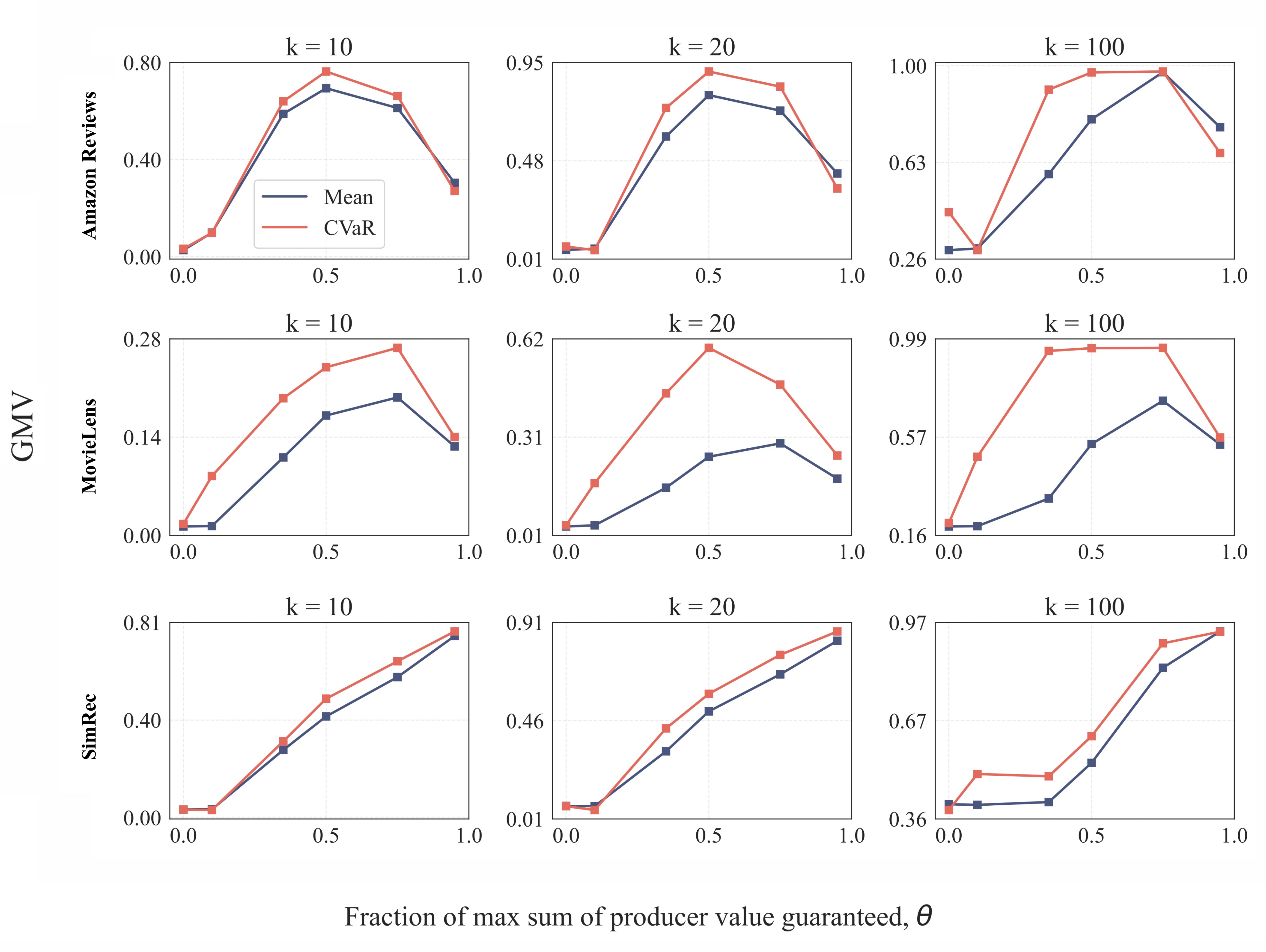}
\caption{Simulated GMV as a function of business-value constraint $\theta$ using mean and CVaR consumer utility objectives.}
\label{fig:sim-gmv}
\end{figure}

 For STR, stricter producer fairness ($\gamma$) diversifies exposure and reduces early sell-outs, often raising STR (Figure~\ref{fig:sim-str}). CVaR further boosts STR by balancing allocations across groups, especially in heterogeneous datasets like SimRec. For GMV, we incorporate producer values $v_j$ inversely proportional to popularity, enforcing $\sum_j v_j \sum_i w_{ij} \geq \theta V_{\max}^{\mathrm{GMV}}$. We stress that this inverse-popularity value signal is a deliberate \emph{assumption} used to simulate a business objective, not a model of real price dynamics: by construction it assigns higher value to less-exposed producers, so the observed GMV gains from fairness should be read as a property of this synthetic value model rather than as direct evidence about revenue in a deployed marketplace. Results in Figure~\ref{fig:sim-gmv} show that moderate thresholds $\theta$ increase GMV by encouraging exposure to high-value producers, while excessive thresholds hurt relevance and reduce transactions. Again, CVaR consistently yields higher GMV than mean optimization, particularly in skewed datasets like MovieLens. \\

\begin{table}[h!]
\caption[Optimization methods for fair allocation on Amazon Reviews.]{
\label{tab:perf}
Optimization methods for fair consumer–producer allocation on Amazon Reviews ($\gamma = 0.5$, $\alpha = 0.95$, $N_{\mathcal{C}} = N_{\mathcal{P}} = 500$). We compare the MIP solver (\textit{Baseline}), relaxed LP (\textit{Rel}), augmented Lagrangian (\textit{AugLag}), and soft-constrained gradient (\textit{SCGrad}), each with hard-threshold (\textit{Round}), probabilistic (\textit{Prob}), or top-$k$ (\textit{TopK}) binarization. We report normalized consumer utility and group utility variance (mean $\pm$ standard error over three runs).
}
\resizebox{0.8\linewidth}{!}{
\begin{tabular}{@{}clll@{}}
\toprule
\multicolumn{1}{l}{\textbf{$k$}} & \textbf{Method} & \textbf{Consumer Util.}     & \textbf{Group Util. Var.} \\ 
\midrule
\multirow{10}{*}{10}                         & Baseline        & $0.95 \pm \scriptstyle{0.00}$ & $0.01 \pm \scriptstyle{0.00}$      \\
                                             & Rel Round        & $0.95 \pm \scriptstyle{0.00}$ & $0.01 \pm \scriptstyle{0.00}$      \\
                                             & Rel TopK         & $0.95 \pm \scriptstyle{0.00}$ & $0.01 \pm \scriptstyle{0.00}$      \\
                                             & Rel Prob         & $0.95 \pm \scriptstyle{0.00}$ & $0.01 \pm \scriptstyle{0.00}$      \\
                                             & AugLag Round    & $0.95 \pm \scriptstyle{0.00}$ & $0.01 \pm \scriptstyle{0.00}$      \\
                                             & AugLag Prob     & $0.95 \pm \scriptstyle{0.00}$ & $0.01 \pm \scriptstyle{0.00}$      \\
                                             & AugLag TopK     & $0.95 \pm \scriptstyle{0.00}$ & $0.01 \pm \scriptstyle{0.00}$      \\
                                             & SCGrad Round    & $0.92 \pm \scriptstyle{0.00}$ & $0.01 \pm \scriptstyle{0.00}$      \\
                                             & SCGrad TopK     & $0.92 \pm \scriptstyle{0.00}$ & $0.01 \pm \scriptstyle{0.00}$      \\
                                             & SCGrad Prob     & $0.92 \pm \scriptstyle{0.00}$ & $0.01 \pm \scriptstyle{0.00}$      \\
                                            \midrule
\multirow{10}{*}{25}                         & Baseline        & $0.94 \pm \scriptstyle{0.00}$ & $0.06 \pm \scriptstyle{0.00}$      \\
                                             & Rel Round        & $0.94 \pm \scriptstyle{0.00}$ & $0.06 \pm \scriptstyle{0.00}$      \\
                                             & Rel TopK         & $0.94 \pm \scriptstyle{0.00}$ & $0.06 \pm \scriptstyle{0.00}$      \\
                                             & Rel Prob         & $0.94 \pm \scriptstyle{0.00}$ & $0.06 \pm \scriptstyle{0.00}$      \\
                                             & AugLag Round    & $0.94 \pm \scriptstyle{0.00}$ & $0.06 \pm \scriptstyle{0.00}$      \\
                                             & AugLag TopK     & $0.94 \pm \scriptstyle{0.00}$ & $0.06 \pm \scriptstyle{0.00}$      \\
                                             & AugLag Prob     & $0.94 \pm \scriptstyle{0.00}$ & $0.06 \pm \scriptstyle{0.00}$      \\
                                             & SCGrad TopK     & $0.90 \pm \scriptstyle{0.01}$ & $0.06 \pm \scriptstyle{0.00}$      \\
                                             & SCGrad Round    & $0.90 \pm \scriptstyle{0.01}$ & $0.06 \pm \scriptstyle{0.00}$      \\
                                             & SCGrad Prob     & $0.90 \pm \scriptstyle{0.01}$ & $0.06 \pm \scriptstyle{0.00}$      \\ \bottomrule 
\end{tabular}}

\end{table}

\noindent \textbf{Scalability and Approximation Methods.} While binary MIPs capture fairness exactly, they scale poorly for large $\pmb{\rho}$. Relaxed LP with rounding and gradient-based methods (AugLag, SCGrad) provide tractable alternatives. Table~\ref{tab:perf} shows that LP relaxation and AugLag match MIP in fairness and mean utility, whereas SCGrad underperforms on utility. Table~\ref{tab:errs} confirms constraint satisfaction is generally reliable, except under probabilistic rounding.

\begin{figure}[t]
\centering
\includegraphics[width=0.75\columnwidth]{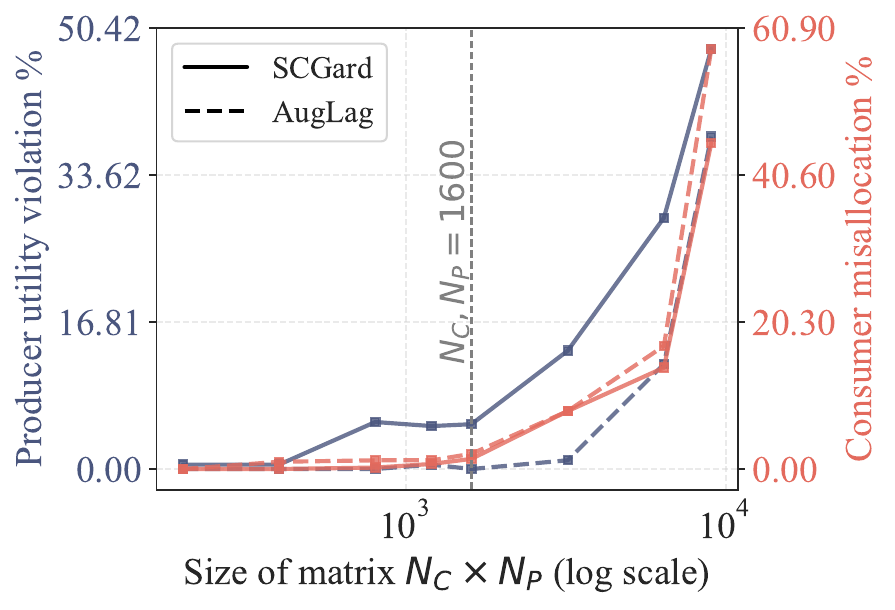}
\caption{Constraint violations for AugLag and SCGrad as a function of relevance matrix size on the Amazon Reviews dataset ($\gamma = 0.5$, $\alpha = 0.95$, $k = 10$). We track the share of consumers with incorrect $k$-allocations and producers below required utility.}
\label{fig:errors}
\end{figure}

 Runtime experiments (Figure~\ref{fig:times}) highlight the trade-off: Gurobi is efficient for moderate scales, but GPU-accelerated methods achieve $1.5\times$ speedups at larger sizes. Constraint satisfaction remains a challenge for gradient-based methods (Figure~\ref{fig:errors}), motivating future work on tighter regularizers.

\section{Conclusion}

 This work formalizes two-sided fairness under realistic heterogeneity and challenges two common assumptions in recommendation system design. The first is that producer fairness is costly. While our results confirm that stricter exposure guarantees reduce average consumer utility—with drops of 15 - 25\% at $k=10$ under full fairness—they also reveal that moderate fairness constraints ($\gamma \approx 0.3$–$0.6$) can \textit{increase} business metrics such as Sell-Through Rate and Gross Merchandise Value by diversifying exposure away from saturated producers. The trade-off is real, but not zero - sum.

 The second assumption is that consumer fairness is adequately captured by aggregate objectives. We show that optimizing for mean or max-min individual utility leaves substantial variance across consumer groups, with disadvantaged segments receiving 20 - 40\% lower relevance than advantaged ones. CVaR optimization compresses this distribution, improving worst-group outcomes without meaningful degradation elsewhere. For platforms serving heterogeneous populations, group-level objectives are not optional refinements but necessary safeguards.

\begin{table}[]
\caption[Constraint violations across optimization methods on Amazon Reviews.]{
\label{tab:errs}
Constraint violations across optimization methods for fair consumer–producer allocation on Amazon Reviews ($\gamma=0.5$, $\alpha=0.95$, $N_{\mathcal{C}}=N_{\mathcal{P}}=500$). Methods and binarization strategies are as in Table~\ref{tab:perf}. We report the percentage of consumers with under-allocation ($\sum_{j}w_{ij} < k$), over-allocation ($\sum_{j}w_{ij} > k$), and producer-utility violations ($\sum_{i}w_{ij} < \gamma\,U_{\min}^{*\mathcal{P}}$).
}
\resizebox{0.9\linewidth}{!}{
\begin{tabular}{@{}cllll@{}}
\toprule
\multicolumn{1}{l}{\textbf{$k$}} & \textbf{Method} & \textbf{Under Alloc. \%}      & \textbf{Over Alloc. \%}      & \textbf{Prod. Utility Viol. \%} \\ \midrule
\multirow{10}{*}{10}                                                & Baseline        & $0.00 \pm \scriptstyle{0.00}$ & $0.00 \pm \scriptstyle{0.00}$ & $0.00 \pm \scriptstyle{0.00}$   \\
                                                                    & Rel Round        & $0.00 \pm \scriptstyle{0.00}$ & $0.00 \pm \scriptstyle{0.00}$ & $0.00 \pm \scriptstyle{0.00}$   \\
                                                                    & Rel TopK         & $0.00 \pm \scriptstyle{0.00}$ & $0.00 \pm \scriptstyle{0.00}$ & $0.00 \pm \scriptstyle{0.00}$   \\
                                                                    & Rel Prob         & $1.00 \pm \scriptstyle{1.00}$ & $1.00 \pm \scriptstyle{1.00}$ & $0.00 \pm \scriptstyle{0.00}$   \\
                                                                    & AugLag Round    & $1.00 \pm \scriptstyle{1.00}$ & $1.00 \pm \scriptstyle{0.00}$ & $1.00 \pm \scriptstyle{1.00}$   \\
                                                                    & AugLag TopK     & $0.00 \pm \scriptstyle{0.00}$ & $0.00 \pm \scriptstyle{0.00}$ & $1.00 \pm \scriptstyle{0.00}$   \\
                                                                    & AugLag Prob     & $6.00 \pm \scriptstyle{1.00}$ & $4.00 \pm \scriptstyle{0.00}$ & $7.00 \pm \scriptstyle{3.00}$   \\
                                                                    & SCGrad Round    & $0.00 \pm \scriptstyle{0.00}$ & $1.00 \pm \scriptstyle{0.00}$ & $0.00 \pm \scriptstyle{0.00}$   \\
                                                                    & SCGrad TopK     & $0.00 \pm \scriptstyle{0.00}$ & $0.00 \pm \scriptstyle{0.00}$ & $0.00 \pm \scriptstyle{0.00}$   \\
                                                                    & SCGrad Prob     & $0.00 \pm \scriptstyle{0.00}$ & $1.00 \pm \scriptstyle{0.00}$ & $0.00 \pm \scriptstyle{0.00}$   \\
                                                                    \midrule
\multirow{10}{*}{25}                                                & Baseline        & $0.00 \pm \scriptstyle{0.00}$ & $0.00 \pm \scriptstyle{0.00}$ & $0.00 \pm \scriptstyle{0.00}$   \\
                                                                    & Rel Round        & $0.00 \pm \scriptstyle{0.00}$ & $0.00 \pm \scriptstyle{0.00}$ & $0.00 \pm \scriptstyle{0.00}$   \\
                                                                    & Rel TopK         & $0.00 \pm \scriptstyle{0.00}$ & $0.00 \pm \scriptstyle{0.00}$ & $0.00 \pm \scriptstyle{0.00}$   \\
                                                                    & Rel Prob         & $1.00 \pm \scriptstyle{0.00}$ & $0.00 \pm \scriptstyle{0.00}$ & $1.00 \pm \scriptstyle{0.00}$   \\
                                                                    & AugLag Round    & $1.00 \pm \scriptstyle{1.00}$ & $0.00 \pm \scriptstyle{1.00}$ & $1.00 \pm \scriptstyle{1.00}$   \\
                                                                    & AugLag TopK     & $0.00 \pm \scriptstyle{0.00}$ & $0.00 \pm \scriptstyle{0.00}$ & $1.00 \pm \scriptstyle{1.00}$   \\
                                                                    & AugLag Prob     & $5.00 \pm \scriptstyle{0.03}$ & $4.00 \pm \scriptstyle{1.00}$ & $5.00 \pm \scriptstyle{2.00}$   \\
                                                                    & SCGrad TopK     & $0.00 \pm \scriptstyle{0.00}$ & $0.00 \pm \scriptstyle{0.00}$ & $0.00 \pm \scriptstyle{0.00}$   \\
                                                                    & SCGrad Round    & $0.00 \pm \scriptstyle{1.00}$ & $0.00 \pm \scriptstyle{0.00}$ & $0.00 \pm \scriptstyle{0.00}$   \\
                                                                    & SCGrad Prob     & $0.00 \pm \scriptstyle{0.00}$ & $0.00 \pm \scriptstyle{0.00}$ & $0.00 \pm \scriptstyle{0.00}$   \\ \bottomrule 
\end{tabular}}

\end{table}

 Methodologically, we extend two-sided fairness from soft single-item allocations to discrete multi-item recommendations, the setting that many practical platforms more closely resemble. We show that the ``free fairness'' phenomenon—where producer constraints impose no consumer cost—disappears in this regime. We use CVaR as a tractable consumer-side objective and integrate business constraints directly into the optimization rather than treating them as post-hoc diagnostics. The exact, LP-based, and gradient-based solvers should therefore be read as tools for instantiating and studying the framework, not as a claim that a single global optimization layer is immediately deployable in every production recommender. \\

\noindent  However, several limitations warrant caution. Our evaluation relies on offline simulations with fixed group structure; real markets evolve, and group definitions are contested. We ignore positional bias, assuming all slots in a recommendation list receive equal attention - an assumption that breaks down for ranked feeds. Our GMV signal is a deliberate synthetic assumption—producer value set inversely proportional to popularity—used to simulate a business objective rather than observed from transactional logs or reflective of real price dynamics; sensitivity analyses under alternative value constructions would strengthen the business-facing conclusions. We also do not benchmark against the broader set of existing fair re-ranking baselines. Our evaluation further centers on CVaR as the consumer-side fairness objective and on a single group definition per dataset; alternative fairness metrics and group definitions would help establish how broadly the observations generalize. Finally, the datasets we use are relatively small and dense (MovieLens-100k and a $10\text{k}\times10\text{k}$ Amazon Reviews subset), so our scalability claims are best read as relative to exact MIP at the tested sizes; evaluating on larger subsets or more market-native datasets (e.g., larger Amazon Reviews samples or Google Local Business data) is an important next step. Gradient-based solvers also occasionally violate constraints at scale, suggesting that tighter regularization or hybrid methods are needed before deployment.

 The central open question this work raises is whether fairness interventions generalize across market conditions. Our experiments span three datasets with different consumer-producer ratios and relevance distributions, but the optimal fairness utility operating point varies across settings. Developing adaptive methods that tune fairness parameters to market dynamics, rather than requiring manual specification, is a promising direction. Equally important is closing the loop with live experimentation: the causal effects of fairness interventions on long - term marketplace health remain unmeasured.

\begin{acks}
The authors gratefully acknowledge the support of Vinted and the University of Amsterdam, without which this work would not have been possible. We are especially grateful to Maarten de Rijke and the Information Retrieval Lab (IRLab) at the University of Amsterdam for their encouragement and guidance.
\end{acks}

\bibliographystyle{ACM-Reference-Format}
\balance
\bibliography{main}

\appendix
\section{Experimental Setup}

\noindent \textbf{Synthetic Dataset Details.}
SimRec simulates $1000$ consumers and $1000$ producers with group heterogeneity. Consumers are partitioned into $G=10$ groups drawn from a Zipf distribution with exponent $1.0$, producing a long-tailed distribution. Each consumer's relevance profile is generated using a monotonically decreasing utility curve over items with additive Gaussian noise:
\[
\rho_i = \text{clip}\left(1 - \frac{\log(1+\beta_g x)}{\log(1+\beta_g)} + \epsilon, 0.1, 1\right),
\quad \epsilon \sim \mathcal{N}(0,0.2^2),
\]
where $x$ ranges uniformly over items and $\epsilon$ is Gaussian noise with standard deviation $\sigma = 0.2$. The smoothness is controlled by a group-specific $\beta$ parameter that decreases with group index, making higher-ranked groups more peaked in their preferences.

\begin{figure}[tbh]
\centering
\begin{subfigure}[b]{0.45\columnwidth}
\centering
\includegraphics[width=\textwidth]{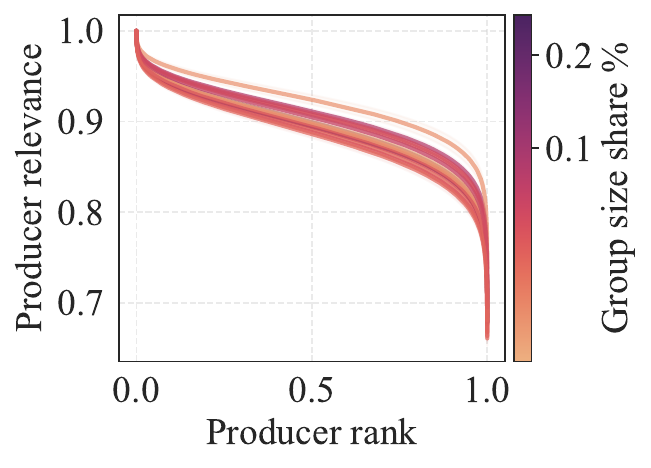}
\caption{Amazon Reviews}
\end{subfigure}
\hspace{0.02\columnwidth}
\begin{subfigure}[b]{0.45\columnwidth}
\centering
\includegraphics[width=\textwidth]{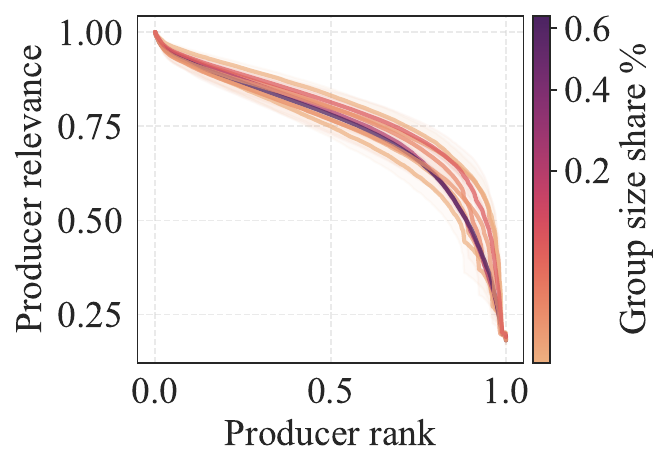}
\caption{MovieLens}
\end{subfigure}

\vspace{1.5em}

\begin{subfigure}[tbh]{0.45\columnwidth}
\centering
\includegraphics[width=\textwidth]{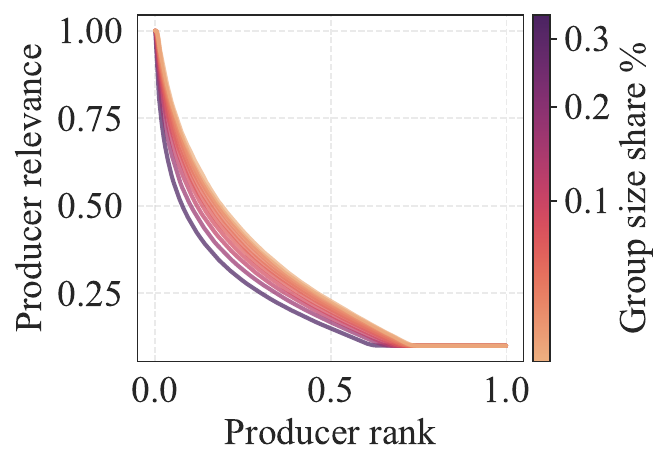}
\caption{SimRec}
\end{subfigure}
\caption{Average producer relevance decay across consumer groups for each dataset.}
\label{fig:rel_fallof}
\end{figure}

\noindent Figure~\ref{fig:rel_fallof} illustrates how producer relevance decays across consumer groups for each dataset. Each curve represents one consumer group, with color indicating group size (darker indicates larger groups). The x-axis shows normalized producer rank (from most to least relevant), while the y-axis shows mean relevance per-group. \\

\noindent \textbf{Recommender System Performance.}
For both MovieLens-100k and Amazon Reviews, we train a two-tower neural recommender model on user-item interaction data. The model achieves high top-$k$ quality, with Precision@10 and Recall@10 scores shown in Table~\ref{tab:rec-res-appendix}.

\begin{figure}[t]
\begin{subfigure}[t]{\columnwidth}
\includegraphics[width=\textwidth]{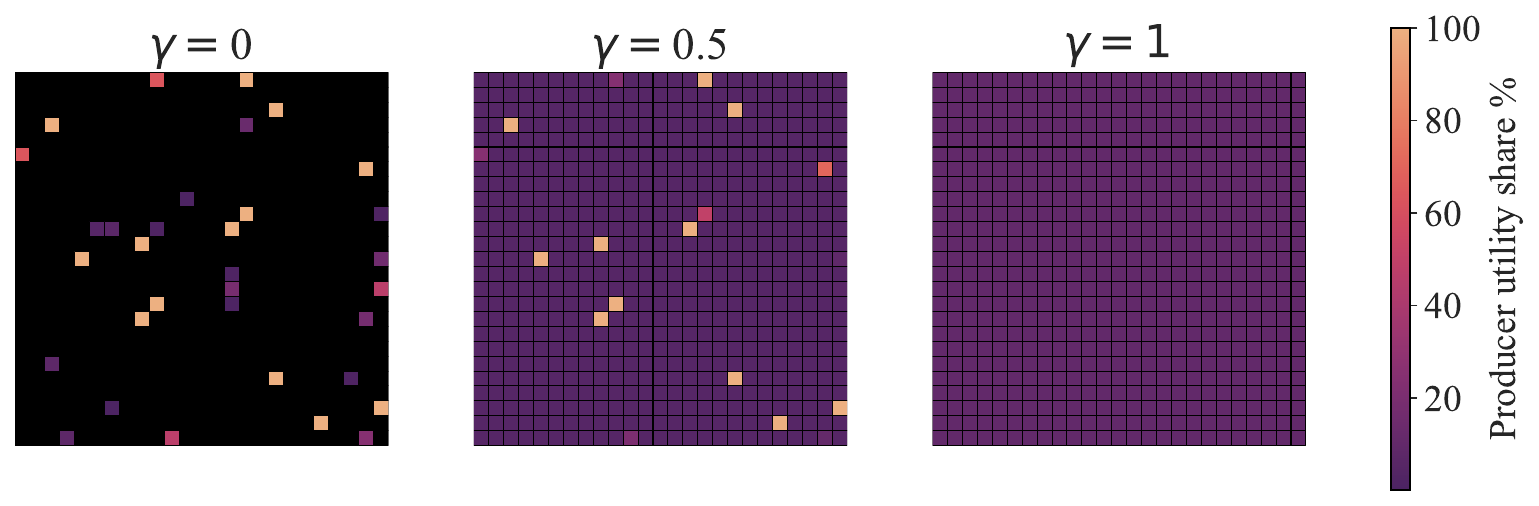}
\caption{Varying producer fairness $\gamma \in \{0, 0.5, 1\}$}
\label{fig:all-per-prod-gamma}
\end{subfigure}
\hfill
\begin{subfigure}[t]{\columnwidth}
\centering
\includegraphics[width=\textwidth]{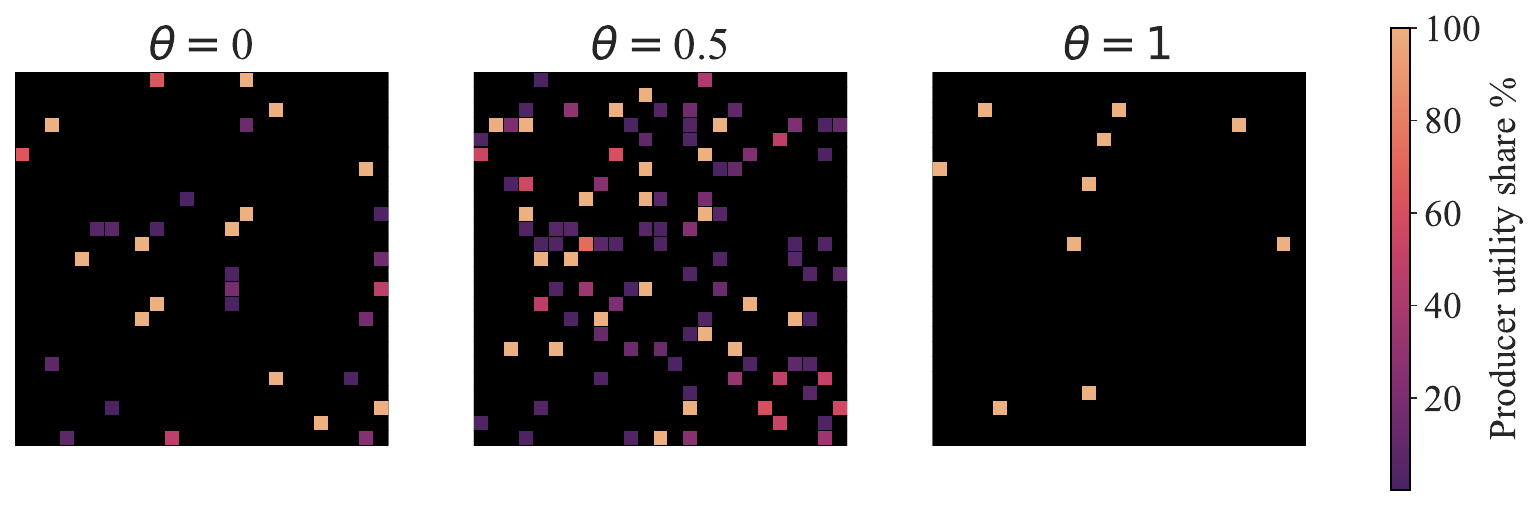}
\caption{Varying GMV threshold $\theta \in \{0, 0.5, 1\}$}
\label{fig:all-per-prod-theta}
\end{subfigure}
\caption{Producer allocation patterns on Amazon Reviews ($k=10$) under (a) different producer fairness levels $\gamma$ and (b) different GMV thresholds $\theta$. Black cells indicate producers with zero allocations.}
\label{fig:allocation-heatmaps}
\end{figure}

\begin{table}[t]
\centering
\caption{Precision@10 and Recall@10 scores of the trained two-tower recommender.}
\begin{tabular}{@{}lll@{}}
\toprule
\textbf{Dataset} & \textbf{Precision@10} & \textbf{Recall@10} \\ \midrule
MovieLens        & 97.41                 & 16.72              \\
Amazon Reviews   & 97.73                 & 23.82              \\ \bottomrule
\end{tabular}
\label{tab:rec-res-appendix}
\end{table}

\section{Optimization Formulations}\label{sec:optimization-frameworks}

\noindent \textbf{Augmented Lagrangian Method.}
The AugLag method reformulates the constrained optimization problem using dual variables and quadratic penalties. Let $\boldsymbol{c}_i=\sum_j w_{ij}$ be the number of items allocated to consumer $i$, and $\boldsymbol{p}_j=\sum_i w_{ij}$ the exposure of producer $j$. With target cardinality $k$ and minimum exposure $p_{\min}$, we define:
\begin{equation*}
\mathcal{L}_{\mathrm{card}}(\pmb{w}) = \sum_{i=1}^m (\boldsymbol{c}_i-k)^2, \; \; \mathcal{L}_{\mathrm{prod}}(\pmb{w}) = \sum_{j=1}^n \bigl[\max(0,\,p_{\min}-\boldsymbol{p}_j)\bigr]^2.
\end{equation*}
The full loss is then given as the combination of losses
\begin{equation*}
\begin{split}
    \mathcal{L}_{\mathrm{aug}}(\pmb{w}) &=
\mathcal{L}_{\mathrm{util}} \\ &+
\boldsymbol{\alpha}^{\top}(\boldsymbol{c}-k\mathbf{1}_m) +
\boldsymbol{\beta}^{\top}\max(0,\,p_{\min}\mathbf{1}_n-\boldsymbol{p}) \\ &+
\tfrac{\lambda}{2}\mathcal{L}_{\mathrm{card}} +
\tfrac{\lambda}{2}\mathcal{L}_{\mathrm{prod}},
\end{split}
\end{equation*}
where $\boldsymbol{\alpha},\boldsymbol{\beta}$ are dual variables and $\lambda$ controls penalty strength. \\

\noindent \textbf{Soft-Constrained Gradient Method.}
The SCGrad method treats constraints as soft penalties, omitting dual variables entirely. We combine the CVaR loss with cardinality and producer terms, and add a binarization regularizer:
\[
\mathcal{L}_{\mathrm{bin}}(\boldsymbol{w}) = \sum_{i=1}^{m} \sum_{j=1}^{n} \left[ w_{ij}(1 - w_{ij}) \right]^2,
\]
which pushes allocations toward $\{0,1\}$. The final loss is:
\[
\mathcal{L}_{\mathrm{total}}(\boldsymbol{w}) =
\mathcal{L}_{\mathrm{util}} +
\mathcal{L}_{\mathrm{card}} +
\mathcal{L}_{\mathrm{prod}} +
\mathcal{L}_{\mathrm{bin}}.
\]
To improve numerical stability, allocations are modeled with a temperature-controlled sigmoid:
\[
w_{ij} = \sigma\left( \frac{z_{ij}}{\eta^{(t)}} \right), \quad
\eta^{(t)} = \max\left( \eta_0 \cdot \text{anneal\_rate}^t, \eta_{\min} \right).
\]

\section{Business Metrics Simulation Details}

We simulate purchases in a stepwise process: 

\begin{itemize}[leftmargin=15pt, topsep=0.4em, itemsep=0.4em]

\item[(1)] For each consumer, compute a weighted score for each producer based on the product of the allocation matrix and the consumer-producer relevance matrix.

\item[(2)] Select the top-$k$ producers with the highest scores.

\item[(3)] For each of the top-$k$ producers, sample a purchase decision using a Bernoulli draw with success probability equal to its relevance score.

\item[(4)] If a purchase occurs, register the transaction and mark the selected producer as sold out, removing it from future allocations.

\end{itemize}

This simulation mimics a real-world transactional setting, where items are recommended, selected based on interest (modeled as relevance), and then become unavailable once sold. It allows us to compare downstream metrics such as total sales, sell-through rate (STR), or average transaction value across different allocation strategies.

Figure~\ref{fig:allocation-heatmaps} visualizes how producer allocations change under different fairness ($\gamma$) and business ($\theta$) constraints on the Amazon dataset with $k=10$. Each square represents a producer, with black squares indicating zero allocations. In the top row, at $\gamma = 0$, most producers receive no exposure, leading to overconcentration; at higher $\gamma$ values, allocations become more diverse.

\end{document}